\def\etal{{\it et~al.~}}
\def\Lya{{\rm Ly}\kern 0.1em$\alpha$}
\def\Mg{{\rm Mg}\kern 0.1em{\sc II}}
\def\MgII{{\rm Mg}\kern 0.1em{\sc II}~$\lambda\lambda2976, 2803$}
\def\C{{\rm C}\kern 0.1em{\sc IV}}
\def\CIV{C\kern 0.1em{\sc IV}~$\lambda\lambda1548, 1550$}
\def\HI{{\rm H}\kern 0.1em{\sc I}}

\def\cm2{\hbox{cm$^{-2}$}}
\documentstyle[11pt,aaspp4,flushrt]{article}
\begin{document}

%\received{date month year}
%\accepted{date month year}
%\journalid{number}{date month year}
%\articleid{number}{number}

\slugcomment{\it submitted to Astrophysical Journal}
\lefthead{Charlton, Anninos, Norman, \& Zhang}
\righthead{Double Lines of Sight}

%%%%%%%%%%%%%%%%%%%%%%%%%%%%%%%%%%%%%%%%%%%%%%%%%%%%%%%%%%%%%%%%%%%%%%%%%%%%%%%

\title{Probing Ly$\alpha$ Absorbers in Cosmological Simulations
       with Double Lines of Sight}

\author{\sc Jane~C.~Charlton\altaffilmark{1}}
\affil{Astronomy and Astrophysics Department \\
       Pennsylvania State University \\
       University Park, PA 16802 \\
       charlton@astro.psu.edu}

\and
\author{\sc Peter Anninos, Yu Zhang\altaffilmark{2}
            and Michael L. Norman\altaffilmark{2}}
\affil{Laboratory for Computational Astrophysics \\
       National Center for Supercomputing Applications \\
       University of Illinois at Urbana-Champaign \\
       405 N. Matthews Ave., Urbana, IL 61801 \\
       panninos, yzhang, norman@ncsa.uiuc.edu}

\altaffiltext{1}{Center for Gravitational Physics and Geometry,
                 Pennsylvania State University}
\altaffiltext{2}{Astronomy Department,
                 University of Illinois}

%%%%%%%%%%%%%%%%%%%%%%%%%%%%%%%%%%%%%%%%%%%%%%%%%%%%%%%%%%%%%%%%%%%%%%%%%%%%%%%
\pagestyle{empty}

\begin{abstract}
We perform a double line of sight (DLOS) analysis of the {\Lya} forest
structures that form and evolve in cosmological N--body/hydrodynamic
simulations.
Pairs of simulated spectra, extracted from lines of sight separated
by distances from $D=12.5$~kpc up to 800~kpc, and a ``control sample''
of unrelated lines of sight, are analyzed at redshifts
3, 2, and 1.  Coincident line samples are defined for {\HI} column density
thresholds $N_{co} = 10^{12.5}$, $10^{13}$, and $10^{14}$~cm$^{-2}$.
We find that: 1) Under the assumption of a single structure size,
a Bayesian analysis yields sizes that are larger for smaller
$N_{co}$, and at fixed $N_{co}$ the size decreases with decreasing
redshift.  However, these derived sizes are found to increase
with increasing $D$ indicating that the assumption
of a single structure size is invalid.
2) The column densities of coincident pairs
are highly correlated for small $D$, with increasing scatter as $D$
is increased, consistent with structures that have a centrally
peaked $N_H$ that decreases gradually with radius.
3) The velocity difference distribution for coincident lines
is very narrow for small $D$, and widens as $D$ is increased
to meet the expectation for chance coincidences in unrelated
lines of sight.  This behavior is indicative of organized motion
within the structures.  4) For small $D$, the distribution
of anticoincident line column densities, $N_{ac}$, falls
steeply as $N_{ac}$ increases from the cutoff value, but
has a significant tail at large values which is
inconsistent with a population of spherical absorbers
with sharp edges, and consistent with a flattened geometry.
The conclusions reached on the basis of the DLOS analysis are validated
by an examination of the three--dimensional structures and velocity
flows in the simulation data.

\end{abstract}


%%%%%%%%%%%%%%%%%%%%%%%%%%%%%%%%%%%%%%%%%%%%%%%%%%%%%%%%%%%%%%%%%%%%%%%%%%%%%%%
\pagestyle{myheadings}
\markboth{\sc Charlton, Anninos, Zhang \& Norman \hfill Double Lines of
Sight~~}
         {\sc Charlton, Anninos, Zhang \& Norman \hfill Double Lines of
Sight~~}

\section{Introduction}

Observations of the spectra of double line of sight (DLOS)
quasars are the most promising method for determining the
sizes and distribution of material in the {\Lya} absorbing
structures and for studying their evolution over time.
Analysis based on present observations is limited by the
number of close quasar pairs that are bright enough
to obtain spectra at high resolution, but this situation
will be rectified as new quasars are discovered.
Numerical cosmological simulations have been quite
successful in matching the properties of observed
structures, even down to the scale of the {\Lya} forest
(\cite{cen94,zha95,her96,mir96}).  In this paper we use
the results of these simulations as a laboratory
in which to conduct DLOS analysis.  The objectives
of this study are to obtain a more intuitive understanding
of the formation and evolution of structure in the
simulations themselves and to aid interpretation of more
detailed DLOS analysis tools in view of the promise of abundant
new data.

Existing DLOS observations demonstrate that the structures
responsible for {\Lya} forest absorption lines are much
larger than predicted by standard CDM mini-halo
(\cite{ree86,mir93}) or pressure confined
(\cite{sar80,ike86}) models.  The strongest constraint
at redshift $\sim 2$ comes from the separation
$\sim 40$~$h^{-1}$~kpc DLOS Q1343+266 A,B (\cite{bec94,din94}).
A Bayesian statistical analysis of
the number of coincident and anticoincident lines
in these spectra yielded a median radius (in the probability
distribution) of $149$~$h^{-1}$~kpc
and a 99\% confidence lower limit of $R = 61$~$h^{-1}$~kpc
(for $\Omega_0 = 1$ where $h = H_0/100$~km~s$^{-1}$~Mpc$^{-1}$)
(\cite {fan96}).  At the lower median redshift of $0.7$,
analysis of the separation $\sim 330$~$h^{-1}$~kpc DLOS
Q0107-0232/0235 (\cite{din95}) gives
an even larger median radius of $501$~$h^{-1}$~kpc
(\cite{fan96}) under the assumption of a single population
of spherical clouds.  However, analysis of several additional
widely separated pairs (\cite{fan96}) yields a trend for an
increasing derived cloud radius with increasing DLOS separation,
even at a fixed redshift, calling into question the simple
single population assumption.  Fang \etal (1995) suggest that a
distribution of sizes and/or correlated (but smaller) structures could
be more consistent. Additional data on DLOS pairs at various
separations and for lines with different column density
cutoffs will be important to determine the evolution of
structure sizes.

Observations of very close QSO pairs (less than tens of kpc)
show nearly identical equivalent widths and redshifts for common
lines (\cite{sme92,sme95,bec96}).   There is some evidence
that the radial velocity difference distribution for coincident
lines widens with increasing DLOS separation, and that the
equivalent widths deviate more from the correlation line
(\cite{din94,fan96}).  These trends would result from
structures with column densities that peak towards the
center and smoothly decrease with radius, and in which some
type of coherent velocity structure is also present
(\cite{ccl95}; hereafter CCL95).  Such structures are
like what one would expect for outer
extensions of galaxies, and in fact at low redshifts
many {\Lya} forest lines appear to result from these
regions (\cite{lan95}).

Here we will perform a detailed DLOS analysis of the
results from two different numerical simulations
(box sizes 3.2 and 9.6Mpc) using our
hierarchical N--body/hydrodynamics code (HERCULES) (\cite
{ann94,ann96}).  Our DLOS analysis will include a Bayesian
analysis of structure sizes, and an application of
the tests developed in CCL95.  That paper presented
the expected distributions of coincident and anticoincident
line properties for ``toy'' models in spherical and disk
geometries.  Three diagnostic tests for geometry and for
the distribution of mass within structures
were applied: 1) comparison of
neutral Hydrogen column densities $N_H$ for lines coincident in
the two lines of sight; 2) distributions of the differences
in velocities between coincident lines; 3) the distribution
of $N_H$ for anticoincident lines.  Application of these tests
to the limited existing data at redshift two (weighted heavily
by consideration of DLOS Q1343+266 A,B) led to the
conclusion that slab/disk--like, smooth structures with
$N_H \sim R^{-4}$ and with coherent internal motion
(such as rotation or inflow) are consistent.

In the study of the {\Lya} forest, the relationship between numerical
simulation data and observational data is symbiotic.  By performing 
DLOS analysis on numerical simulations we gain the ability
to probe all different separations over the full
range of redshifts.  This enables a diagnosis of
the nature of the {\Lya} structures produced by
the simulations and an exploration of their evolution.
The results of this DLOS analysis will be compared
to those from similar analysis of existing observational
data.  Finally, the simulation data have
the distinct advantage of providing complete knowledge of
the distribution of matter in space, velocity, and time.
This provides a check on intuitions developed purely
on the basis of DLOS analysis, and allows
further development of tools that can
be used both for simulations and for observational data.

This paper begins in \S 2 with a summary of our numerical
techniques and methods used to generate and analyze simulated spectra.
Section 3 presents the results of the various DLOS
tests, including a Bayesian size analysis, column density
distributions of coincident lines, velocity differences
of coincident lines, and column density distributions
of anticoincident lines.  We conclude with a discussion in \S 4
of the various lessons learned from the analysis
of simulation data.

%%%%%%%%%%%%%%%%%%%%%%%%%%%%%%%%%%%%%%%%%%%%%%%%%%%%%%%%%%%%%%%%%%%%%%%%%%%%%%%
\newpage
\section{Numerical Methods}
\label{sec:methods}

The cosmological simulations discussed in this paper
were conducted using the three-dimensional
hierarchical code (HERCULES) that we have developed to solve
the $N$-body equations for the
dark matter particles, the hydrodynamic equations for the baryonic gas,
and the cosmological expansion equation for the FLRW background universe.
The code also models the radiative cooling of the gas, supplies an
external flux of ionizing radiation, and solves the
non-equilibrium rate equations for 27 chemical reactions
involving nine atomic and molecular
species (including the various ions of Hydrogen and
Helium).  Further details and tests of this code can be found in
Anninos, Norman \& Clarke (1994) and Anninos \etal (1996).

We have performed two simulations using different box sizes
with comoving dimensions 3.2 Mpc and 9.6 Mpc.
The smaller 3.2 Mpc box calculation is initialized with a cold dark matter
(CDM)
fluctuation spectrum normalized to $\sigma_8 = 1$, with a Hubble
constant $H_0 = 70$ km s$^{-1}$ Mpc$^{-1}$, total density parameter
$\Omega_0 =1$, and baryon fraction $\Omega_B = 0.04$. The baryonic component is
composed of Hydrogen and Helium in cosmic abundance with a Hydrogen mass
fraction of 76\%.
A photoionizing radiation field with spectral index
$\alpha=1.5$ is turned on at $z=6.5$ and increased
in redshift to a maximum value of
$J_{912} = 8\times 10^{-23}$ ergs s$^{-1}$ cm$^{-2}$ Hz$^{-1}$ sr$^{-1}$
at $z=4.5$, held constant until $z=0.5$, then decreased linearly with redshift
to 10\% of $J_{912}$ at $z=0$.
The parameters of the larger 9.6 Mpc box calculation are similarly defined
except
that a second year COBE normalization of the CDM fluctuations
is used with $\sigma_8 = 1.36$.  Also, for the large box
simulation we set $H_0 = 50$ km s$^{-1}$ Mpc$^{-1}$ and $\Omega_B = 0.06$.
The radiation is increased with redshift from $z=6.5$ to a larger
amplitude (than the 3.2 Mpc case)
$J_{912} = 2\times 10^{-22}$ ergs s$^{-1}$ cm$^{-2}$ Hz$^{-1}$ sr$^{-1}$
at $z=4.5$, held constant, then decreased linearly from $z=2$
to 10\% of $J_{912}$ at $z=0$.

Taken together, the two simulations allow us to simultaneously explore
the effect of varying the physical parameters,
grid resolution, and large
scale gravitational tidal fields.
In particular we note that
the larger box simulation has a cell size of only
$\Delta x = 19$ kpc at $z=3$ (and a factor of two worse at $z=1$), and
therefore will not be of value for comparisons
of close lines of sight.  The small box simulation has
a resolution of 6.3~kpc at $z=3$, and thus we can consider
DLOS with separations 25~kpc, and with some caution, even
12.5~kpc.  In order to best examine the formation
and clustering of structures (that can be several hundred
kpcs in size) it is important to consider the effect of
the surrounding environment and the distribution of matter on
large scales, thus the large box simulation
is complementary to the small box simulation which has higher
resolution and greater ability to resolve cooling flows.

We have performed a DLOS analysis on the results of both the
3.2 Mpc and 9.6 Mpc box simulations.
Spectra were generated for various random lines of sight
through the simulation boxes using a convolution procedure
(\cite{bi95,zha96}).  The optical depth is computed at
each point in velocity space by integrating
$\tau = c\int n_{H} \sigma dt$ (where $n_{H}$ is the
number density of neutral Hydrogen, $c$ is the speed of
light, $dt$ is the interval of time,
and $\sigma$ is the absorption cross section assuming
a Voigt profile for the lines) along a LOS from the QSO source.
The total velocity for an absorption feature is the combination
of the Hubble flow velocity and the peculiar velocity, which
is computed from the local gravitational acceleration,
pressure gradient forces, and bulk fluid motion for each contributing cell.
The transmission along the line of sight is then given by $e^{-\tau}$.

A series of parallel lines of sight for the 3.2 Mpc box at $z=2$
are shown in Fig. 1.  Qualitatively, the two spectra are
nearly identical (both in the presence of lines coincident
in redshift and in the strengths of these lines) for
small separations, begin to differ substantially
as the DLOS separation increases, but still exhibit
similarities even on the scale of hundreds of kpc.
The spectra also appear similar over different redshift ranges except
that the opacities become smaller with decreasing redshift at a rate
that depends on the cosmological expansion, the ionizing radiation
flux, and the density of lines. The density of lines itself
decreases along the DLOS with decreasing redshift,
in agreement with observations (\cite{lu91,bec94,bah96}).

To facilitate a quantitative analysis of the DLOS
we must identify and extract individual lines from the spectra.
Absorption features are identified as locations in
the spectra where the opacity is greater than some pre--determined critical
value, typically 0.2. A second selection procedure is then applied to
identify individual lines as those features which have a clear
central maximum opacity.
These lines are then fit with Gaussian profiles, allowing for multiple
profiles to provide adequate fits to saturated and/or blended features.
The Doppler parameters and column densities of the extracted lines
are computed from the line center opacity, the equivalent width
$W_\lambda = \int \left(1-e^{-\tau}\right)d\lambda$, and the curve of growth
for a Maxwellian velocity distribution (\cite{spit78}). Further details
of this procedure are provided in Zhang \etal (1996).

For each of three
redshifts ($z = 3$, 2, and 1) and two box sizes ($L = 3.2$ and 9.6~Mpc),
and for seven DLOS
separations ($D = 12.5$, 25, 50, 100, 200, 400, and 800~kpc),
we produce 1000 pairs of parallel spectra, each extending
over a redshift interval $\Delta z = 0.1$.
In the remainder of this paper, we will focus on what can be learned from
the DLOS analysis of these simulated data.

%%%%%%%%%%%%%%%%%%%%%%%%%%%%%%%%%%%%%%%%%%%%%%%%%%%%%%%%%%%%%%%%%%%%%%%%%%%%%%%

\section{Double Line of Sight Analysis}
\label{sec:analysis}

In analyses of observational data, the sizes of {\Lya}
absorbing structures were estimated by using
the numbers of ``coincident'' and ``anticoincident''
lines in DLOS (\cite{din94,bec94,din95,fan96}).
A coincidence was defined as a case where lines
were present in both lines of sight above some
$S/N$ threshold and
within a velocity interval of $\Delta V < 150$~km/s.
An anti-coincidence was defined as a case of a line
in one LOS without a detected line in the
other LOS within $\Delta V < 150$~km/s.  In the
rare case that two lines in one line of sight fall
within $150$~km/s of a line in the other line of
sight, Fang \etal (1995) counted only one coincidence,
but also did not include the third line as an
anticoincidence.  For the equivalent width detection thresholds of
these spectra (using $\le 4$~m telescopes) the probability
of coincidences occurring by chance is minimal.
However, for more abundant weaker lines (as in these
simulated data or in forthcoming 8--m class telescope
spectra) and/or at high redshift
this will no longer be the case.

In view of the difficulties of correcting for random
coincidences, we adopt the following simplest procedure
for defining coincident pairs.  For each line in one
LOS we define a coincidence with the
closest line in velocity along the other LOS if
this line is within our coincidence velocity
criterion $\Delta V_{co}$ (either $150$~km/s or
$50$~km/s).  An anticoincidence is counted only
if there is no line within $\Delta V_{co}$ in
the other LOS.  This leaves us with another set of
lines that are within $\Delta V_{co}$ of
another line, but are not the closest in
velocity space, which we shall call adjacent
lines.

For each redshift and box size, and for two choices of
$\Delta V_{co}$ we analyze separately the
set of lines exceeding column density thresholds $N_{co} =$
$10^{12.5}$, $10^{13}$, and $10^{14}$~cm$^{-2}$.
Fig. 2 displays for the 3.2~Mpc box simulation
(solid curves)
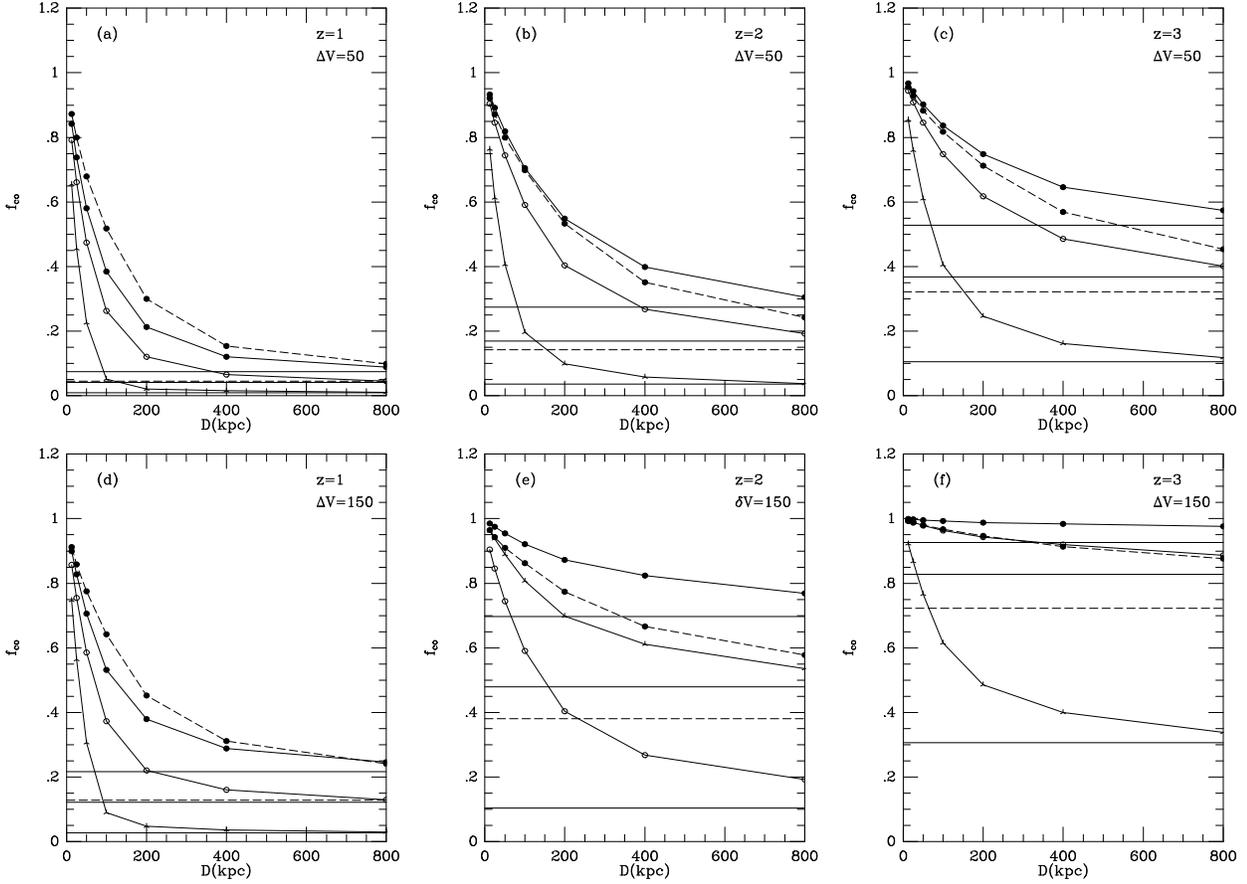
the ratio of the number of coincidences to the sum of the
number of coincidences and anticoincidences,
$f_{co}$, as a function of the DLOS separation.
The fraction of coincidences generally decreases with increasing
$D$, but levels off at large $D$.
The slow flattening of these curves at large separations is
attributed to chance coincidences.
For $\Delta V_{co} = 150$km/s, nearly all lines are
coincident if $z \ge 2$ and/or $N_H \le 10^{13}$~cm$^{-2}$
The more restrictive cutoff,
$\Delta V_{co} = 50$~km/s,
is generally more useful for our analysis.

To consider the effect of chance coincidences on DLOS
analyses we construct ``control samples'' of unrelated
lines of sight.  This is accomplished by dividing each
of the 1000 lines of sight (for a given redshift and box
size) at the midpoint into two equal parts and analyzing
each of these pairs as if it was a DLOS.  The fractions
of coincident lines to arise in these ``control samples'',
for the three $N_{co}$ values in the 3.2~Mpc box,
are given as solid horizontal
lines in Fig. 2.
The fraction of chance coincidences
becomes comparable to the fraction of coincidences for real
DLOS at large $D$, but there is an ``excess'' of real
coincidences at small $D$ in all cases.  Clearly,
chance coincidences make a significant contribution
for large $z$ and small $N_{co}$.

The dashed curves and horizontal lines in Fig. 2 give a
comparison for $N_{co} = 10^{12.5}$~cm$^{-2}$ in the 9.6~Mpc box simulation.
Several effects compete to determine the effect of the box size on $f_{co}$.
The 9.6~Mpc box has a more realistic
representation of large scale structure, and a larger cell size, which
together act to increase $f_{co}$ over both large and small $D$.
However, because of the higher spatial resolution in the 3.2~Mpc box, the
density of lines along the spectrum is greater than in the larger box.
The smaller box thus has a larger fraction of chance coincidences
(compare the dashed and the upper solid lines), which increases
$f_{co}$ at large $D$. This latter effect dominates for $z \ge 2$,
so that the 3.2~Mpc box simulations have larger $f_{co}$.
For the cases with $z=1$, the levels of chance coincidences in
both the 3.2~Mpc and in the 9.6~Mpc simulations are small
compared to $f_{co}$, particularly at small $D$, and the
effect of large scale power in the larger box leads to
a larger $f_{co}$.

It is not straight--forward to correct for the effect of chance
coincidences, but we can estimate their relative importance.
Subtracting the estimated number of chance coincidences leads
to overcorrection: in cases with
large $f_{co}$ it is clear that many of the
coincidences are indicative of real structures so that
the actual number of chance coincidences is much
smaller than the number derived from unrelated LOS.
As an example, consider a sample of all lines with
neutral column density $N_H > 10^{12.5}$~cm$^{-2}$.
For the small box simulation at $z=2$, with the largest
DLOS separation of
$D = 800$~kpc, and with $\Delta V_{co} = 150$~km/s, in our
1000 DLOS
there are $45200$ coincidences, 13644 anticoincidences,
and 13877 adjacent lines.  If we are more restrictive, with
$\Delta V_{co} = 50$~km/s, the number of coincidences is
reduced to 24393, the number of anticoincidences
increases to 55379, and the number of adjacent lines
decreased to only 3313 (3\% of the total number of lines).
If all coincidences were
by chance, since the fraction of the lines in one LOS
that are coincident is 0.45 we would expect a fraction $(0.45)^2$,
or 20\% of the lines to be adjacent. The small observed
number of adjacent lines is indicative that a significant
number of the coincidences are real, even at large $D$.
In the following analysis we will not attempt to
quantitatively ``correct'' for chance coincidences,
but will refer to Fig. 2
to assess the implications of this effect.

\subsection{Maximum Likelihood Sizes}

With the simplified assumption that the structures
responsible for {\Lya} absorption (above a given $N_{co}$)
are a single population of uniform size
we can apply Bayesian statistics to
estimate the structure sizes at the various $N_H$
contour levels.  We follow the procedure given in
Fang \etal (1995), using the numbers of coincidences
(or ``hits'' ${\cal N}_h$) and anticoincidences (or ``misses'' ${\cal N}_m$)
to evaluate the
probability $P(R)$ that the radius of the
single population has value $R$, for a spherical
and for a disk geometry.  As in Fang \etal (1995),
we define $X=D/2R$ so that the probabilities, for
the two geometries, that a second LOS intersects
the structure if a first one does are:
$$
\phi_{sphere} = \frac{2}{\pi} \left [ \cos^{-1} X - X (1 - X^2)^{1/2} \right ]
        \qquad \mbox{for} \quad X<1~,
$$
$$
\phi_{disk} = \int_{-\pi/2}^{\pi/2}
              \frac{\cos\theta}{\pi} \left
[\cos^{-1}\left(\frac{X}{\cos\theta}\right)
              - \frac{X}{\cos\theta}\left(1 -
\frac{X^2}{\cos^2\theta}\right)^{1/2}
              \right ]~d\theta \qquad \mbox{for} \quad X<\cos\theta~,
$$
and $0$ otherwise.
Defining the related quantity, $\psi = \phi/(2-\phi)$, the
probability that both LOS intersect the structure, knowing that
at least one of them does,
we can calculate the maximum likelihood probability that
the structure has radius $R$ as
$$
P(R) = {{({\cal N}_h+{\cal N}_m+1)!}\over{{\cal N}_h! {\cal N}_m!}}
\psi^{{\cal N}_h} (1-\psi)^{{\cal N}_m} \frac{d\psi}{dR}
$$
where
$$
\frac{d\psi}{dR} = \frac{8}{\pi} \frac{X}{R} (1-X^2)^{1/2} (2-\phi)^{-2}.
$$

Fang \etal ``correct'' for
random coincidences by subtracting the number of chance
pairs expected (aside from those that would be
classified as adjacent lines) from the observed
number and by adding twice this number to the number
of anticoincidences.  This procedure works well in
cases where the density of lines is low (high $N_H$ threshold
or low $z$), but it is not sufficient if the number of
accidental hits is comparable to the number of
coincidences even for very close lines of sight.
For our analysis, we simply used the observed number of
coincidences and anticoincidences, disregarding adjacent
lines, and acknowledging
that the resulting sizes will be overestimated,
particularly for large DLOS separations.

Figure 3 presents the median sizes from $P(R)$ derived
from the Bayesian
analysis of various sets of the simulated spectra as a function of DLOS
separation.  If the assumption of a single, uncorrelated
population of structures is valid and if contamination by
chance coincidences is insignificant, we expect the size derived
to be constant at various DLOS separations.  Yet we see that
for all curves the derived size increases with increasing
$D$.  Each panel of Fig. 3 gives three curves, corresponding
to $z = 3$, $2$, and $1$, and the series on the top
shows the median sizes at the three contour levels
$10^{12.5}$, $10^{13}$, and $10^{14}$~cm$^{-2}$ derived with the
criterion $\Delta V < 50$~km/s for coincident lines (for the
large 9.6 Mpc box simulation).
Smaller sizes result from the assumption of a spherical
cloud population, as illustrated by a comparison of Figs. 3a
and 3d.  The effect of the box size/resolution is
illustrated by a comparison of the median radii
for $10^{13}$~cm$^{-2}$ contour levels in
Figs. 3b (9.6~Mpc box) and 3e (3.2~Mpc box).
For this case, derived sizes are somewhat larger
in the 9.6~Mpc box, because the effect of enhanced
power on larger scales dominates over the effect
of a larger chance pair contribution in the small box.
Finally, the effect
of relaxing the $\Delta V$ criterion to $150$~km/s is shown
in Fig. 3f for the $10^{14}$~cm$^{-2}$ contour.  This less restrictive
criterion can add more ``real'' coincidences as well as
more accidental ones, but at large $D$ most of the increase
of the derived median size is likely to be due to the added
contamination.

The assumption of single sized structures is only an approximation,
but it is still useful to give rough numbers for the sizes of
structures at the different $N_{co}$ and $z$.
The most reliable size estimate will come from
an intermediate $D$ range (roughly
comparable to the structure size).  Smaller $D$ estimates are
suspect due to the low resolution of the simulations on these
scales, while at larger $D$ the chance pairs make a larger
contribution to the measured number of coincidences.
The median size does not change too rapidly in the regime
where $D \sim R$.  We derive the rough numbers from the
large box simulations because of the more realistic handling
of large scale power, and because of the lesser contamination
by chance coincidences (due to the smaller line density).
Ultimately, a better method is needed to quantify
the size for cases with a high density of lines.

The following basic results for sizes and size evolution
can be extracted from the simulation data:
1) At all redshifts we find that the sizes get larger
with decreasing $N_{co}$.  In the next section we will
evaluate whether this is a consequence of the higher $N_H$
regions being embedded in the lower $N_H$ regions.
2) As the redshift decreases, the median sizes of structures
decrease. This effect is not as pronounced at large $N_{co}$
as it is at small $N_{co}$.  Roughly, at $10^{12.5}$~cm$^{-2}$
the median radius of the disk/sphere
decreases from 1200/800~kpc
at $z=3$, to 600/400~kpc at $z=2$, to 300/200~kpc at $z=1$.
At $10^{13}$~cm$^{-2}$ the median disk/sphere radii are $\sim$1000/600~kpc
($z=3$),
500/300~kpc ($z=2$), and 250/200~kpc ($z=1$).
At the $10^{14}$~cm$^{-2}$ contour level the median radii are
$\sim$300/200~kpc ($z=3$), 175/100~kpc ($z=2$), and 125/100~kpc ($z=1$).
These sizes must
be interpreted with caution because accidental coincidences
are more likely at high $z$ and they do tend to push upward
the size estimates, however they are based on the
large sized box where the effect of chance coincidences
is less important.
3) The size constraints increase with increasing DLOS
separation.  This is likely to be partially due to
contamination by chance pairs, but
it is also of interest to note that this agrees with the
result of Fang \etal (1995) in their similar analysis of observational
data.  They point out that this effect could be a result
of a non-uniform size population or of correlated structures.
4) Our size estimates from the numerical simulations
(100-300 kpc at $z=2$ over the
$N_{co}$ range $10^{13}$ --- $10^{14}$~cm$^{-2}$) are quite
consistent with the median of $R = 150$~kpc derived from
the observations of Q1343+266 A,B assuming a population of spheres.

\subsection{Column Densities for Coincident Lines}

In Fig. 1 of CCL95 we presented results of Monte--Carlo
simulations that generate the distributions of column densities
for pairs of coincident lines in several idealized cases
in which {\Lya} absorption is produced by a single
population of identical structures.  We considered an
irregular distribution of material for which column densities
are unrelated to each other, independent of DLOS separation,
and we considered spherical and disk structures in which
the column density falls off gradually from a central peak.
In all cases we assumed that the one population of structures
has an $N_H$ distribution consistent with the observed distribution
$f(N_H) \sim N_H^{-1.5}$ for the entire
population of {\Lya} forest lines.  For ``smooth sphere'' or
``smooth disk'' structures this power law $f(N_H)$ results
from a column density law $N_H(R) \sim R^{-4}$.
In the CCL95 toy models, for a completely irregular
distribution there is no correlation between DLOS column
densities for any separation $d = D/R$.  However, for the
smooth structure, Fig. 1 in CCL95 shows highly correlated
column densities $N_A$ and $N_B$ for small $d$, increasing
spread as $d$ is increased, and then a tendency to have
small $N_A$ for all large $N_B$ (and vice versa) for
$d \sim 1$.

Here, in Fig. 4 we show $N_A$ vs. $N_B$ for
all coincidences at $z=2$ in the 3.2Mpc box with $N_H > 10^{12.5}$~cm$^{-2}$,
defined by the $\Delta V_{co} < 50$~km/s criterion.
The sequence a--e shows, with increasing $D$,
the increasing spread in the correlation between $N_A$ and $N_B$.
For the large box simulation, the correlation at small $D$ is considerably
tighter between $N_A$ and $N_B$ because
12.5~kpc is sub-cell so that A and B
are not sampling independent locations.
However, it is illustrated in Fig. 4f that the correlation
is stronger for the 9.6~Mpc box even when $D=100$~kpc,
and this is likely to be a real effect related to the enhanced
large-scale power, since 100~kpc is
safely larger than the resolution ($\sim 25$ kpc) at $z=2$.

The results of Fig. 4 appear very similar to
the smooth density distribution toy models in Fig. 1
of CCL95, indicating that such a model is a fairly
good description of the numerical simulation results.
The spread in this series of plots with increasing $D$
can be used to match them to the toy models, which
are expressed in terms of the dimensionless parameter
$d=D/R$.  Roughly, the $D=50$~kpc appears most similar to the
toy model with $d=0.1$.  This comparison
yields an independent estimate of the structure size:
$R \sim 500$~kpc at the $10^{12.5}$~cm$^{-2}$ contour level.
At the largest $D$ displayed (= 200 kpc, Fig. 4e), the
distribution of $N_A$ vs. $N_B$
resembles that for an irregular distribution, with the
column densities unrelated between the DLOS.  As
$D$ is further increased we do not observe in the
simulations the tendency for large $N_A$ values when
we have small $N_B$ values, as seen for the toy models.
This is not surprising, since at large $D$
many ``coincidences'' occur just by chance,
and since this effect will be washed out by a
more realistic distribution of structure sizes
and shapes.

For higher column density thresholds, the $N_A$ vs. $N_B$ plots
appear qualitatively similar, however the ``match'' to the toy
models occurs for a smaller $R$ value.  At $z=2$, for coincidences with a
$10^{13}$~cm$^{-2}$ threshold, the $D=50$~kpc figure has scatter consistent
with $d \sim 0.25$ in the toy models, yielding
$R \sim 200$~kpc.
Similarly, for a $10^{14}$~cm$^{-2}$
threshold we find that $d=0.5$ corresponds to $D=50$~kpc so
that $R$ is about 100~kpc.

The sizes estimated here are fairly consistent
with the Bayesian analysis in the previous section,
based on the numbers of coincidences and anticoincidences,
certainly within the 90\% confidence limits.
It is interesting to consider the implications of these results
for the density of material
within the structures.  The consistency with the toy model
indicates a smoothly decreasing distribution of mass around density peaks,
but not necessarily a single population of separate structures.
The sizes estimated in this section relate to the rate of
decrease of $N_H$ with $R$ around a density peak.  This
will be discussed further in the concluding section.

A related DLOS plot is the equivalent width difference between
coincident lines in A and B vs. the maximum equivalent width
of the two lines.  For the $D=40$~kpc DLOS Q1343+266 A,B at
$z \sim 2$, Fang \etal (1995) find a trend for high equivalent width
lines to have larger equivalent width differences.
In the simulations, a series of plots with increasing
$D$ would show points moving from the horizontal line
with zero equivalent width difference to the correlation
line representing the maximum equivalent width difference.
Fig. 5 presents the results from the simulation
with box size 3.2Mpc at $z=2$, with $N_{co} = 10^{14}$~cm$^{-2}$
and $\Delta V_{co} = 150$~km/s,
the case that should most closely resemble the 1343+266 A,B
observations which have an equivalent width cutoff of about
0.3\AA.  The simulation results appear consistent
with the limited data for Q1343+266 A,B
observations, but note that the simulations predict
a scatter of $|W_A - W_B|$ values at large max$(W_A,W_B)$
that would only become apparent in a larger observed sample.

Finally, the $N_A$ vs. $N_B$ plots provide
another way of considering the level of contamination
by chance coincidences.  These chance coincidences should
be distributed in the $N_A$ vs. $N_B$ diagram as an
irregular toy model distribution would be, or
equivalently as a distribution with a very large $D$ value such as
in Fig. 4e.  If these chance pairs are
a dominant contributor to the total number of coincidences
we should see a significant scatter from the correlation
lines.  We see some outlying points, in Figs. 4a
and 4b for example, that could be due to chance pairs, but
this is not a significant effect.  Such contamination
is observed to be more pronounced at $z=3$ (and less pronounced
in the 9.6~Mpc box simulations), but it is significant that
a correlation between $N_A$ and $N_B$ is still quite apparent
at $D=100$~kpc even in the $z=3$ simulation results.

\subsection{Velocity Differences for Coincident Lines}

Next, we examine the distribution of velocity differences
between coincident lines for the various DLOS separations.
Figs. 6a, b, and c show (for $z=3$, 2, and 1) the distributions
for all coincident lines with $N_{co} = 10^{12.5}$~cm$^{-2}$
in the small 3.2 Mpc box, chosen for its
high resolution at small $D$ separations.  We display all
the data up to $\Delta V = 150$~km/s, but many of the contributing
pairs at $\Delta V > 50$~km/s are chance coincidences (as
discussed above) so we
should focus on comparisons at small $\Delta V$.  These distributions
are normalized so that $f(\Delta V) = 1$ at 5~km/s (the
center of the first bin).  The
results are nearly identical in the 9.6 Mpc simulations.
For very small
$D$ nearly all coincident lines have $\Delta V < 10$~km/s,
and the distribution gradually widens with increasing $D$.
There is an apparent difference between the distributions
at the three different redshifts for a fixed $D$ value,
but in interpreting this trend we must again consider
the effect of contamination by chance pairs.
At $D=800$~kpc the $f(\Delta V)$ distributions for coincident
pairs are nearly identical to
those derived from chance coincidences in the ``control''
sample (filled circles).  These large $D$/control sample distributions
would be flat except for the omission of adjacent lines
from the coincident sample, an effect that is most
significant at large redshifts.  At $z=1$ where chance
coincidences have a minor contribution, the $D=800$~kpc
case is nearly flat.  At large $D$ where the contribution
of chance coincidences is significant, the evolution
of the width of the $f(\Delta V)$ distribution from $z=3$
to $z=1$ has a large contribution from contamination,
and is therefore difficult to interpret.  For smaller
$D$, the effect of contamination (see Fig. 4)
is small, as evidenced by the narrow $\Delta V$ distribution.

The sets of velocity distribution curves in Fig. 6 closely
resemble those expected for a disk with systematic motion,
either rotation or inflow/outflow in the plane (see Fig. 3
in CCL95).  If we use the comparison to estimate $d$ values,
as we did in the previous section for the $N_A$ vs. $N_B$
curves, we obtain similar results.  For example, at $z=2$
the curve with $D=200$~kpc approximately matches the
$d = 0.5$ curve for a disk with inflow/outflow velocity of
$\sim 100$~km/s, yielding $R \sim 400$~kpc for the
$10^{12.5}$~cm$^{-2}$ contounr level.  In CCL95,
we noted a difference expected between rotation and inflow/outflow
in a plane, with a predicted decrease in $\Delta V$ distribution
width for $d>1$ in the case of inflow/outflow.  Such a trend
would be difficult to recognize here due to the uncertainties
of chance coincidence contributions.

The $\Delta V$ distribution
can be compared at the different redshifts if the curves
are labeled by their appropriate values of $d$.  For Figs.
6a, b, and c the $D=800$~kpc curves correspond
to $d = 1$, $2$, and 4, respectively.  With this rescaling,
the velocity distributions for a given $d$
are narrower at lower redshifts.
For coincident pairs with the larger column density cutoffs,
$N_{co}$, we find that the $\Delta V$ distributions are wider
at a given $D$.  This is shown for $N_{co} = 10^{14}$~cm$^{-2}$
at $z=3$ in Fig. 6d.  However, if the
curves are labeled with estimates of the appropriate
values of $d=D/R$ from Fig. 3 then the
series of curves are similar for the different $N_{co}$
values.  The shapes of these histograms are consistent
with models of rotating disks or of inflow/outflow in
a sheet.  If this type of systematic motion is responsible,
then a narrowing of the distribution with time would
represent a decrease in the rotation or inflow/outflow
velocity.

\subsection{Anticoincident Column Density Distributions}

The anticoincident column density distribution was considered
in CCL95 as a method to distinguish between spherical and
disk geometries.  For a smoothly falling density profile
in a sphere, small $D$ DLOS should always yield relatively small values
of the anticoincident column density, while for a disk of arbitrary orientation
it is possible to obtain a larger value.  As the DLOS separation
is increased, we expect to obtain the distribution
$f(N_{ac}) \sim N_{ac}^{-1.5}$.  The number of anticoincidences
is decreased by the presence of chance coincidences, but
there should be no bias with $N_H$, i.e. the remaining
anticoincidences should have the same $f(N_H)$ as the
true full distribution.

In Fig. 7, the anticoincident
column density distributions are presented
for $N_{co} = 10^{12.5}$~cm$^{-2}$ and $\Delta V_{co} = 50$~km/s and for
$N_{co} = 10^{14}$~cm$^{-2}$ and $\Delta V_{co} = 150$~km/s
(for the 3.2~Mpc box simulations at $z=3$, 2, and 1).
Again, the simulation curves resemble the expectations for a smooth
column density profile (eg. with $N_H(R) \sim R^{-4}$ as
in the CCL95 toy models) but are inconsistent with a spherical geometry.
For a DLOS with $d<0.25$ through
a sphere with a sharp edge, there should not be any anticoincidences
with $N_{ac}$ more than three times the cutoff value.
For $N_{co} = 10^{12.5}$~cm$^{-2}$ at $z=2$, all curves with $D < 200$~kpc
must have $d < 0.25$ (using our previous size estimates)
and we see many anticoincidences with large
$N_{ac}$ values.  However,
all series of curves do show an excess of small $N_{ac}$ values
(relative to large ones)
for close DLOS separations, similar to the expectations for
the disk geometry.  In this geometry the large $N_{ac}$ values
arise from LOS that pass close to the center of a disk that
is highly enough inclined that the other LOS falls beyond the edge.
For $N_{co} = 10^{14}$~cm$^{-2}$, the smallest $D$ value of $12.5$~kpc
corresponds to $d \sim 0.05$ -- $0.12$ over the range of redshifts,
$z=3$ to $z=1$.
The deficit of large column densities, i.e. the steepening of the slope
at small $N_{ac}$ of the curves with small $d$ values, is significant: for the
total number of anti--coincidences in the simulations (thousands
even for $D=12.5$~kpc) these
results are not affected by small number statistics.

We conclude that there is a tendency for anticoincidences
for very small DLOS separations to be biased toward small
column densities, i.e. it is not as common to find a {\it much} larger column
density along a LOS close to another LOS for which $N_H < N_{co}$.
However, the fact that relatively large column densities {\it are}
occasionally observed close to an anticoincident LOS implies
the simulations are not composed of spherical structures with sharp edges.
Rather, the results are again consistent with flattened structures
with a relatively smooth density profile, since some of these can be oriented
so that two LOS separated by small $D$ can have drastically different
column densities.

%%%%%%%%%%%%%%%%%%%%%%%%%%%%%%%%%%%%%%%%%%%%%%%%%%%%%%%%%%%%%%%%%%%%%%%%%%%%%%%

\section{Summary and Discussion}
\label{sec:summary}

We have analysed data from
two cosmological N--body/hydrodynamic simulations
(with box sizes 3.2~Mpc and 9.6~Mpc)
to study the formation and evolution of the {\Lya} forest
in a flat CDM dominated universe.  Synthetic spectra were
generated at $z = 3$, 2, and 1 to mimic the observations of double
quasar lines of sight, and analyzed with various tests
developed to study DLOS as if they were observational data.

{\it We find that a DLOS analysis of the numerical
simulations yields results very similar to those obtained
from idealized models in which {\Lya} absorption is produced
by a single population of disks with $N_H(R) \sim R^{-4}$}
(CCL95).  In particular:
1) Lines coincident in both lines
of sight exhibit an increasing spread of column densities
from the  correlation line as the DLOS separation is increased
(Fig. 4).
2) The velocity difference distribution for coincident
lines gets wider for more widely spaced DLOS (Fig. 6),
as expected for coherent motion in the plane of a disk.
This behavior is also expected for pure Hubble flow, and
with this diagram alone we cannot distinguish the two.
3) The distribution of column densities for anticoincidences
does exhibit some large values for DLOS with all separations,
though at small $D$ there are a smaller fraction of 
large values than in the sample of
all {\Lya} lines, i.e. than in the $f(N_H) \sim N_H^{-1.5}$ distribution
(Fig. 7).  This indicates that spherical clouds
are not a good description of the simulation data.
4) A Bayesian estimate of structure sizes based on the observed
numbers of coincidences and anticoincidences yields
larger sizes for lower $N_H$ contour levels (Fig. 3),
as would result from a density distribution that decreases
smoothly from a central value.

Although the DLOS analysis of the simulation data is
generally suggestive of absorption
arising from a population of similarly sized structures,
with smoothly declining density profiles from a central
maximum, some aspects of the analysis show that this
is a simplification.  The Bayesian analysis of structure
size yielded a size that increases with the DLOS separation
used to probe it.  Interestingly, the same result was
obtained from analysis of observational data at several
different separations by Fang \etal (1995) who find
it to be consistent with some models of different
size distributions and/or clustering of somewhat smaller
absorbing structures.
Our DLOS analysis of simulation data adds the additional
constraint that a correlation between the column densities
of DLOS must hold out to fairly large $D$, thus
there must be fairly large individual structures involved.

The size of structures at fixed column densities
is found to decrease with decreasing redshift
in the simulation data, although this could be partly due to
the effects of contamination by chance pairs and the increased line densities
at high redshifts.  Roughly, we find that derived sizes
for structures defined
by the $N_{co} = 10^{12.5}$~cm$^{-2}$ contour level decrease from
$\sim$1000~kpc to a couple hundred kpc from $z=3$ to
$z=1$, while at the $N_{co} = 10^{14}$~cm$^{-2}$ level they
decrease from a couple of hundred kpc to about 100~kpc
in radius.  These constraints are quite consistent with
the median radius of $149$h$^{-1}$~kpc at $N_{co} \sim 10^{13.5}$~cm$^{-2}$
and $z =1.8$ derived from observations of DLOS Q1343+2640 A,B
(\cite{din94,bec94,fan96}).
When scaled by the structure size (i.e. at a fixed value
of $D/R$, the ratio of the DLOS separation to the structure
radius), the internal structure and motion of the {\Lya}
absorbers, as diagnosed by DLOS analysis, is similar at
different redshifts and $N_H$ contour levels.

All of the conclusions that we have reached are subject
to an uncertainty, which will also affect future DLOS
observations more than it has in the past.  As the
column density threshold for detecting lines is
decreased, and as the redshift is
increased, the probability of detecting ``coincidences''
for completely unrelated lines of sight will increase.
This effect tends in the direction of biasing estimates
toward larger sizes for high redshifts and small $N_H$.
It is simple enough to correct for the problem in the case
where a few such accidental coincidences are expected,
but in the case where the probability of such a false
coincidence approaches unity for each line (as for
$N_{co} = 10^{12.5}$~cm$^{-2}$ and $z=3$) the situation
is more confused.
In view of the fact that 8--m
class telescopes will push DLOS detection thresholds
to considerably lower levels ($N_{co} \sim 10^{12.5}$~cm$^{-2}$), 
new techniques must
be developed to understand the problems associated
with random superpositions of unrelated lines.

This DLOS analysis of the numerical simulation data yields
results consistent with similar analyses of the much smaller
amount of observational data available thus far
(\cite{ccl95,fan96}).  If this were an analysis of real observed
spectra then we would have to accept at face value the
inference that absorption arises from flattened structures
with $N_H$ that fall smoothly from peak values and with
coherent velocity structure.  However, the numerical
simulation laboratory allows us to check these conclusions,
and to examine in more detail the nature of the internal
structure and motion.

Regarding the distribution of material responsible for
absorption, we would like to check and see if a typical
structure really does show a decrease of column density
with radius in the way that our idealized single
population models would require, i.e. $N(R) \sim R^{-4}$.
The following questions are also of interest: To what
extent are the structures disk-shaped?  Are they individual
structures or are they interconnected on a larger scale?
What fraction of the absorption at a given contour level
is produced in ``smooth'' density law structures?
How uniform in size are structures at various $N_H$
contour levels?

To qualitatively address these points, we display
in Fig. 8 contour diagrams for a slice
one cell thick through the 3.2 Mpc box at redshifts 3, 2, and 1
(from left to right).
The top row sequence shows surface levels of neutral Hydrogen
column densities $N_H$ across the single projected cell. The second
row of figures shows the same evolution sequence but for the divergence
of the peculiar gas velocity. In each case, the slice is chosen to
intersect the densest structure in the box.

The structures responsible for absorption can extend through more than
a single cell. In fact,
at $z=3$ contours of $N_H = 10^{12.5}$ cm$^{-2}$ appear as an
extended and interconnected system
of sheet--like structures that surround regions of higher density.
(However, we emphasize that the column densities quoted here are
integrations across only a single cell and are thus likely to be
much lower than those computed from the DLOS integrations across
the whole of these structures.)
Higher contour levels ($10^{13}$ cm$^{-2}$) define more compact
but elongated structures such as filaments. Still higher
column density contours ($> 10^{14}$ cm$^{-2}$) define
spherical knots or halos and are typically found
at the intersections of filaments.
At lower redshifts, the mean density throughout the box decreases with the
expansion of the universe and the $10^{12.5}$~cm$^{-2}$ contours are no longer
simply connected in sheet--like structures, but are more likely to
surround higher density regions such as filaments. They are much less
extended, while the higher density contours are less
noticeably different in size.
The sheet--like structures are still present but are now identified with
lower column densities $\sim 10^{11}$ cm$^{-2}$.
Results from the Bayesian analysis of structure sizes as a function of redshift
are consistent with the picture presented here. The lower density thresholds,
which define extended structures at high redshifts,
correspond to more compact objects as the universe expands to smaller
redshifts, and structures at the same threshold thus appear to
decrease in size.
The higher density structures are less sensitive to this effect as they are
compact to begin with, and their evolution in size is not as pronounced.
It is of interest to note that the decreasing structure size noted
in these simulations is opposite to the conclusion from DLOS size
analysis of observational data (\cite{fan96}).  However, the only
available low redshift DLOS ($z \sim 0.7$) has a large separation
($D \sim 330$h$^{-1}$~kpc) and we have seen in Fig. {sizes}
that size estimates based on large $D$ DLOS are inflated.

Figure 8 also provides a direct estimate for the size of
structures.
For example, at $z=2$, the filamentary structures appear to have a coherence
length ranging from a few hundred kpc up to about 1 Mpc, with a thickness
of 100 --- 200 kpc. The higher density spherical
structures have typical radii up to a few hundred kpc. This
is again consistent with the DLOS analysis.
We also note that by computing the density field around the
spherical structures
we can test the hypothesis that $N_H(R) \sim R^{-\alpha}$, with
$\alpha=4$, is roughly correct in describing the distribution of column
densities. We have done this experiment by integrating the densities
along 1/4 of the box size in the $z$ direction, centered on the highest
density peak. The resulting two-dimensional grid of column densities
are then averaged and binned into a spherical distribution, centered on the
peak. We find that in the central most regions $R\le 10$ kpc the
distribution varies as a power law with index $\alpha\sim 4$,
steepens significantly to a maximum $\alpha\sim 9$ at the intermediate
radius $R\sim 50$ kpc, then becomes shallower again with
$\alpha\sim 3$ at $R\sim 200$ kpc. An averaged effective index over this
entire domain is roughly consistent with the distribution
$N_H(R) \sim R^{-4}$.

The velocity structure in the absorbers can be understood
by examining the contour map of the local divergence in the peculiar
velocity flow (the redshift sequence along the
second row in Fig. 8).
This map shows that the
gas around the high density contours is generally falling in
towards the central core regions,
due to gravitational and cooling instabilities. This is
evidenced by the negative velocity divergence which is correlated
nicely with the high density spherical structures found in the $N_H$ contours.
There is also a component of velocity, due mostly
to gravity, that is responsible for the merging of coherent
matter distributions
within the larger sheet--like structures.
In the low density regions or voids, the velocity divergence
is mostly positive, suggesting that material is moving
away from the the centers of the voids towards the
higher density filaments.
Although the flow is complex, it appears there is coherent motion
in the structures that is consistent with the diagnosis of the
DLOS analysis.

Basically, the DLOS analysis of numerical simulation data,
although incomplete, yielded
fairly reliable insights into the nature of the structure
responsible for {\Lya} absorption.
Further information can be extracted from more detailed examination
of the three-dimensional
maps in position and velocity afforded by the simulations.
Double line of sight analysis of
future observational data will not yield this level
of detail, but it should be possible to utilize the
tests developed here to recognize the basic nature
of the observed structure.

\acknowledgments

We acknowledge support from NASA grant NAGW--3571 at Penn State,
NSF grant ASC--9318185 (funding the Grand Challenge Cosmology
Consortium GC$^3$) and PSC grant AST950004P at NCSA.  The simulations
and data analysis were
performed on the Convex-3880 and the Convex Exemplar at the National Center for
Supercomputing Applications, University of Illinois at
Urbana-Champaign, and the Cray C90 at the Pittsburgh Supercomputing Center.

%%%%%%%%%%%%%%%%%%%%%%%%%%%%%%%%%%%%%%%%%%%%%%%%%%%%%%%%%%%%%%%%%%%%

\end{document}